\documentclass[nonacm, sigconf]{acmart}
\setcopyright{none}
\renewcommand\footnotetextcopyrightpermission[1]{} 
\usepackage{graphicx}

\begin{document}

\title{Personalized Review Ranking for Improving Shopper's Decision Making: A Term Frequency based Approach}


\author{Akhil Sai Peddireddy}
\affiliation{\institution{University of Virginia}}
\email{ap3ub@virginia.edu}


\begin{abstract}

User-generated reviews serve as crucial references in shopper's decision-making process. Moreover, they improve product sales and validate the reputation of the website as a whole. Thus, it becomes important to design reviews ranking methods that help shoppers make informed decisions quickly. However, reviews ranking has its unique challenges. First, there is no relevance labels for reviews. A relevant review for shopper A might not be relevant to shopper B. Second, since shoppers cannot click on reviews, we have no ways of getting relevance feedback. Eventually, reviews ranking suffers from the lack of ground truth due to the variability in the standard of relevance for different users. In this paper, we aim to address the challenges of helping users to find information they might be interested in from the sea of customer reviews. Using the Amazon Customer Reviews Dataset collected and organized by UCSD, we first constructed user profiles based on user's personal web trails, recent shopping history and previous reviews, incorporated user profiles into our ranking algorithm, and assigned higher ranks to reviews that address individual shopper's concerns to the largest extent. Also, we leveraged user profiles to recommend products based on reviews texts. We evaluated our model based on both empirical evaluations and numerical evaluations of review scores. The results from both evaluation methods reveal a significant increase in the quality of top reviews as well as user satisfaction for over 1000 products. Our reviews based recommendation system also suggests that there's a large chance of user viewing and liking the product we recommend. Our work shows the basic steps of developing a ranking method that learns from a particular end-user's preferences. 

\end{abstract}

\maketitle
\section{Introduction \& Motivation}
In the modern e-commerce industry, customer reviews for a product not only significantly influence potential customers' shopping decisions but also exert great impact to the company's credibility. User-generated reviews have the power to gain customer trust and encourage customers to directly interact with the seller.

According to studies conducted by BrightLocal in 2014 and 2019, in the U.S., up to $88\%$ of the respondents trust online reviews as much as they do personal recommendations and $93\%$ of customers spend more than a minute reading reviews \cite{Brightlocal2019}. Another study shows that $92\%$ of consumers hesitate to make a purchase if there are no customer reviews, and $97\%$ shoppers indicate that customer reviews factor into their buying decisions \cite{FanFuel2017}. 

It is also equally important that the reviews are ordered correctly with the most customer pertinent review ranked higher and placed at the top. As per Amazon's own data:
\begin {enumerate}

\item 70\% of customers who shop on Amazon website never click beyond the first page of search results.
\item 35\% of shoppers on Amazon click on the product featured first on the search page
\item 81\% of the user clicks are on brands presented on the first page of search results.

\end {enumerate}

Our motivation for this work stems from the importance of providing personalized review ranking to customers to enhance their shopping experience. It is imperative that reviews that address shoppers' main concerns and needs should have higher priority as even for the same product, different shoppers have different priorities. Hence, we believe that shopping websites should present the results that best fit the customers interests and priorities, in order to improve user experience of shopping on the website. However, there are several challenges that need to be solved before we can generate helpful ranking methods for customer reviews. First of all, unlike traditional document results, reviews do not have an absolute relevance labels. Since different users might focus on different features of a product, the notion of relevance might vary greatly for these users. Moreover, we cannot get feedback from user's clicks. This makes evaluation of review ranking methods hard as we cannot know if the users are satisfied with the ranking of reviews. Also, there is no ground truth for ranking. That being said, we cannot rely on most of the methods we have discussed in class if we want to determine whether a review ranking method is actually producing satisfying results or not.

For this work, we only selected the Mobiles and Accessories category in the Amazon Reviews Dataset to refine our scope. With that being said, our model and general approach can be applied product reviews in all kinds of categories with little or no modifications. We first index all the reviews from the products. A user profile is created and maintained based on every user's past activities including previous posts, browsing and shopping history. Specifically, user profiles are created by collecting reviews text from products user previously engaged with, applying necessary processing such as stemming and stop word removal, and calculating the weighted frequencies of the words where weights are tuned based on user's past activities. As a result, every user profile contains the words this user cares the most. We then use the created user profiles as queries into our ranking model, namely BM25. The ranking of reviews reflects each individual user's preference and expectation about given a product. Eventually, we utilize the user profile to recommend products based on review texts, which are then shown to be more credible source than product description for recommendations.

Since reviews ranking suffers from the lack of ground truth and the variability in the standard of relevance for different users, we employ both empirical and numerical evaluation methods. After reading the reviews ranked by our ranking system, we observe that most of the reviews in lower positions are generally very short and does not contain too much information about the product while the review placed at the top often contain detailed shopping and using experience and the reviewer's opinion on this product. Regarding numerical evaluation, we design a score, cumulative position-weighted score, to reflect user satisfaction. The results show that there is an average of 20\% increase in user satisfaction using our personalized ranking system and the default Amazon ranking system. Beside ranking reviews based on personalized user profiles, we also propose a method of rating products at term level based on user's preferences. For example, if a user cares very much about the price of a product, we'll offer another rating of the product solely based on the price factor. While we did not do a thorough investigation in this area, we believe that this rating method can also be applied to recommendation systems.

The following sections will be dedicated to explaining our work in detail. Sections 2 presents existing works in personalized ranking. Section 3 introduces the dataset we used, and section 4 details the methods we applied in the experiment and gives corresponding proof and reasoning about the choices we made. Section 5 documents the experiments we did and the results we obtained from experiments. Section 6 describes how we evaluated the ranking method we came up with. Section 7 briefly covers the recommendation system we developed. Section 8 points out the limitations of our work and possible future directions that could be further explored.

\section{Existing Work}
Most existing online shopping platforms only support general, simple, but not very helpful ways of ranking customer reviews. For example, Amazon shoppers can only sort reviews based on two criteria -- ``Top Reviews" and ``Most recent". The rank of a review in the ``Top Reviews" ranking system is largely determined by factors including the number of helpfulness votes the review has received, the time the review was posted, and the reviewer's rank. The Amazon reviewer rank is determined by the total number of helpfulness votes the reviewer has received for all his/her reviews, factoring in the number of reviews he/she has written. Also, the more recent the reviewer's reviews are, the higher the reviewer will be ranked. We failed to find out how Amazon integrates the above factors into its final review ranking algorithm.

Nevertheless, while the ranking method mentioned above seems reasonable at first glance, it fails to provide personalized guidance and help for potential customers \cite{BehavioralVar}. A review that some shoppers find helpful because it addresses these people's concerns about this product. However, since different shoppers have different concerns, the review may fail to help other shoppers. Moreover, shoppers behave differently from time to time on web search. The behavioral difference consists of the amount of time and the number of revisits users spend on a particular web page or review, which can be further analyzed to reflect users' interest on a particular item.

There are also some previous works on the personalized web search. One method \cite{BrowsingHistory} represents each individual users' profiles using a list of terms and corresponding weights associated with those terms, based on user's long-term browsing records. The weighted scores will then be considered into the ranking, along with the relevance of each document. Another method \cite{AutoAnalysis} uses BM25, a known probabilistic weighting scheme, to rank documents. Both methods show statistically significant improvement in the interleaving test, compared to the non-personalized web search. 

There also have been researchers focusing on the personalized ranking of user-generated content. Burgess et al. (2013) \cite{Burgess2013} proposed a service, BUTTERWORTH, that finds content more relevant to users' interests on their feeds on Twitter without using explicit user input. The service contains three major components, a lit generator that groups users into different sub-communities, a list labeller that assigns a human-readable topic to each group, and a topic ranker that ranks topics, which are eventually presented to the user. Uysal and Croft (2011) \cite{Uysal2011} designed a personalized ranking of tweets by exploiting retweeting patterns and examining the correlation between retweeting and the interestingness of the tweets for an individual user. They first ranked incoming tweets based on the possibility that users will retweet them. Then they ranked users for each tweet, placing users who are more likely to retweet the tweet at higher positions. Finally, they studied the correlation between ranking method based on retweet likelihood and actual user preferences through pilot user study. While these work provides  helpful insights about personalized ranking of user generated contents, there are significant differences between product reviews and user-generated contents in general. For example, we are unable to get user's feedback through clicks on product reviews. At this point, there is very little work done on exploring the ranking methods for user-generated product reviews specifically.

\section{Dataset Description} 
\subsection{UCSD Amazon review dataset} 
We will use the Amazon Customer Reviews Dataset on UCSD as our main source of data. Due to the limited resources, we  deploy our ranking algorithm on the 5-core dataset of mobiles and accessories category of the entire dataset in the experimental section. The 5-core dataset mentioned above is the subset of the data in which all users and items have at least 5 reviews. 

\smallskip
Format of the 5-core data: 
\begin{description}
  \item reviewerID: "A2SUAM1J3GNN3B", 
  \item asin: "0000013714", 
  \item reviewerName: "J. McDonald", 
  \item helpful: [2, 3], 
  \item reviewText: "I bought this for my husband who loves playing piano.",
  \item overall: 5.0, 
  \item summary: "Heavenly Highway Hymns", 
  \item unixReviewTime: 1252800000, 
  \item reviewTime: "09 13, 2009" 
\end{description}

Some of the terms are explained below:

\begin{itemize}
    \item reviewerID - Unique reviewer identifier
    \item asin - unique Product identifier
    \item reviewerName - Name of the reviewer
    \item helpful - Helpfulness rating of the review, e.g. 2/3
    \item reviewText: Text of the review
    \item overall - Rating given by a user for the product
    \item summary - Summary of the review
    \item unixReviewTime - Time when the review is published (unix time)
    \item reviewTime - Time when the review is published (raw time)
\end{itemize}

\subsection{Exploratory data analysis}
In the mobiles and accessories category of UCSD Amazon reviews dataset, there are totally 194, 439 reviews, 27, 879 unique users, and 10429 products. Based on the 5-data summary table presented below, we can see that the median number of reviews by a user is 7 and the median number of reviews associated with each product in the dataset is 32. From the distributions of different attributes of the data, we can see that the distributions of number of reviews written by each user and the number of reviews associated with each products are highly skewed to the left. Furthermore, the maximum number of reviews for a product is 837. These discoveries lead us to adopt the indexing framework that maps each term to its frequencies in the reviews instead of using the inverted index, as discussed in Section 4. 

\smallskip
\begin{center}
\begin{tabular}{ | c | c | c | c | c |}
 \hline
  & \# of u reviews & \# of p reviews & rating & review length \\
  \hline
  mean & 9.936 & 81.512 & 4.123 & 491.840 \\
  \hline
  std & 13.234 & 130.246 & 1.222 & 749.170 \\
  \hline
  min & 5 & 5 & 1 & 0 \\
  \hline
  25\% & 5 & 13 & 4 & 143 \\
  \hline
  50\% & 7 & 32 & 5 & 248 \\
  \hline
  75\% & 9 &  82 & 5 & 532 \\
  \hline
  max & 152 & 837 & 5 & 32110 \\
  \hline
\end{tabular}
\end{center}

\begin{center}
\textbf{Note}: 
\textit{\# of u reviews} is the number of reviews by each users \\
\textit{\# of p reviews} is the number of reviews associated with each product
\end{center}

\begin{figure}[htp]
\centering
\includegraphics[width=8cm]{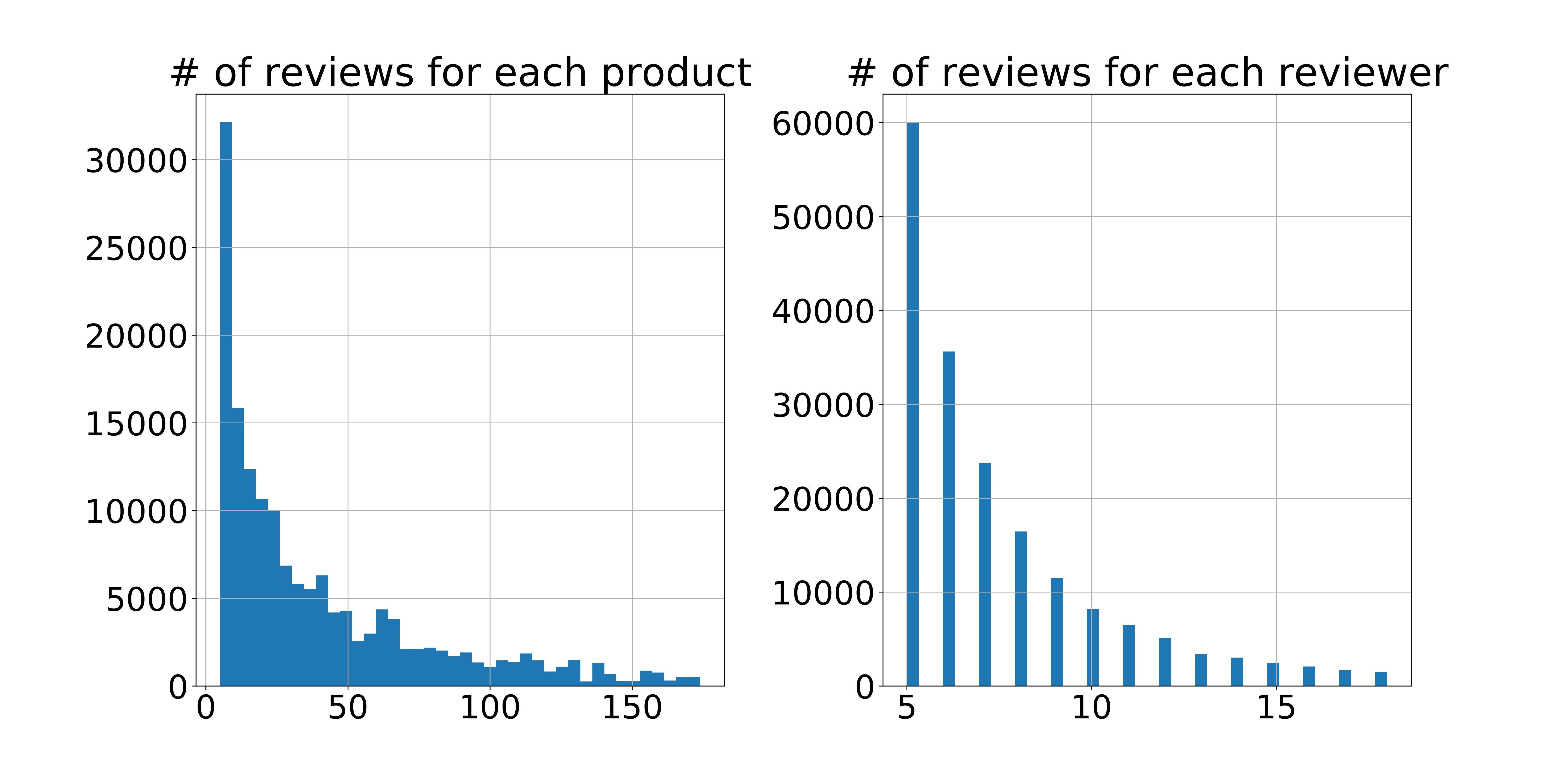}
\caption{The outlier-removed distributions of the number of reviews by each user and the number of reviews for each product}
\end{figure}

\section{Proposed Method}
The goal of Information Retrieval is to increase user satisfaction. To achieve this goal in the optimal review ranking, it is desired to rank the reviews and recommend products on a personalized basis taking into consideration the diverse interests of users. To achieve this task of personalized review ranking, we propose a method which takes the frequency of the terms in the reviews of previously engaged products into consideration, to rank the reviews on the target product. The steps involved in the method are as follows.

\begin{enumerate}
    \item Indexing all the tokenized and pre-processed reviews of the products. The pre-processing steps can include stemming, stopword removal, converting to lower case etc.
    \item Creating and maintaining a user corpus with individual profile from the activity on each shopping category. This user profile has the terms the user cares the most.
    \item Taking user profile as a query which reflects the preferences of the user, rank the reviews on the target product to provide most user's  reviews at the top.
    \item Since Review text is a more credible source than product description for recommendation, leveraging the user profile to recommend products based on their review text.
\end{enumerate}

\subsection{Review Text Frequency Model (RTFM)}
User profile decides the effectiveness of the above suggested method. But how to construct the user profile in a way that truly reflects the preferences of the user ? How to find the set of words the user care the most ? If a user is browsing for Mobiles, is that battery or camera or price or quality that is important to him. We propose a model which we coined as Review Text Frequency Model (RTFM) which can provide a solution the questions we have. The idea of the model is to use the frequency of the terms in reviews to construct the user profile. Let's take a detailed look into how this model works.

\begin{enumerate}
    \item Access the Indexed Reviews for the products user previously engaged with.
    \item Based on the category we are trying to create a user profile, decide if it makes sense to do Parts of Speech tagging and remove terms other than noun, adjective. This step can improve performance on most of the categories.
    \item Calculate weighted frequencies of the words and get the 'top k' words with highest frequency. The weights can be decided based on the type of the activity as each activity has a different importance.\\
    We suggest to use the weights as follows.
    \begin{itemize}
    \item Visited product : Weight based on stay time. Reward for longer time and punish for shorter time.
    \item Shopped product : high weight
    \item Previous reviews of the user : highest weight
    \end{itemize}
Note that this weight is a parameter which can be tuned.i.e, how much 'high' for a shopped / reviewed product.
    \item When the user has a new activity, update the user profile with the terms and their new frequency. The frequency of the term can be updated as,
    \begin{equation}
    freq\ (t) _{\ new} = freq\ (t)_{\ old} + weight * freq\ (t)_{\ product}
    \end{equation}
\end{enumerate}

\subsection{Design Decisions}
\begin{enumerate}
\item Inverted Index vs ‘Not Inverted' Index\\
One of the reasons for using Inverted Index is that is makes search easy while the query is a set of few words and the set of documents to be searched is very huge. Incase of a search engine this number goes to billions of documents. However in our case, as see in exploratory data analysis, the number of reviews per product is less than 1000, and that is generally the trend as well. As the search has to be done only on the reviews of the target product, with a large query of k words, we argue that a forward index is more suitable than an inverted index as the number of reviews on a product / business in most websites are typically far fewer than the words in vocabulary. 
\newline

\item Why choose RTFM ?\\
RTFM captures ‘what people are talking' with respect to a product, irrespective of the sentiment of the popular opinion (both positive and negative considered). As an example, if we consider people talk more about quality, price and camera about an Apple iPhone in it's reviews, RTFM automatically gets this information. This model also implicitly learns the user preferences after a series of user activity. 
\newline

\item Unigram vs n-gram model\\
Two scenarios where n-gram model helps is, for learning sentiment and context. But for review ranking a neutral sentiment is to be maintained. To make a shopping decision, the user wants to see both positive and negative reviews for the terms he or she cares about. If the review is negative, then the user will ideally skip buying that product. Although context can be learned by n-gram model, that's a trade-off one can take for efficiency as the advantage is not high as in other applications of Information Retrieval.
\end{enumerate}

\section{Experiments \& Results}
\subsection{Proof of Concept}
Before moving to the actual experiments on generating user profiles, to make sure the user preferences can convert to desired review ranking order, we manually constructed our user profile based on our preferences and passed them as a query to rank the reviews. We observed the reviews returned at top were quite relevant and helpful to make a decision about buying the product. Here's how our query looks like "reliable, camera, light, simple, lightweight, good, slim,durable, pixel, quality, android, cheap, long, lasting, reception,quality,sturdy, picture, call, signal, safe, investment, value, money, features". The top ranked review is - "Great features-except for the phone one.Seriously-Bluetooth, IR, good phone book features, nice color display.However, I get much weaker signals(and call quality) on this phone.If you are on the edge,this is not the phone for you unless you value the non-phone features more than it working as a phone" and the least ranked review is - "Its simply awesome. What else to say?". The top ranked review is indeed helpful and reflects some of our preferences which proves the assumption that rightly modeling user preferences and using them as a query to rank reviews can produce assist in shopper's decision making. 

\subsection{Constructing Reviews \& User Corpus}
As discussed in the Proposed Method, the first step is to index all the reviews with their terms and the corresponding frequencies after necessary pre-processing steps. The next step is to create a user corpus for each user and for each category. Since we were working on the Mobiles and accessories alone, we had to create a user corpus with all user profiles. But we didn't find any dataset that has the real user browsing, shopping activity so we decided to simulate this data on random for ALL users. The total browsed products by each user is chosen randomly between 100 and 500 and products are also randomly picked. For shopped products, we chose the number randomly between 30 to 100. The previous reviews are directly accessed since we already have that data in the dataset. The weights are each activity, used to update the term frequency is chosen as follows,
\begin{enumerate}
    \item Browsed product - According to Time of browsing\\
    <=1 minute => Weight = -2\\
    1-5 minutes => Between -2 and 2\\
    2.5 minutes => No update as user preference is unclear\\
    >=5 minutes => Weight = 2
    \item Shopped product => Weight = 5
    \item Terms in previous review => Weight = 10\\
    Then we calculated the weights for all the terms using the formula in Section 4.1.
\end{enumerate}
We now have a huge user corpus with profiles of all the users where each user profile consists of all the terms along with their weighted frequencies. We pick the 'top k' words and in our case we selected k as 300. Now when the user visits a new product, we pick the tokenized reviews of the product from the 'reviews corpus' and the user profile of the user from the 'user corpus'.

\subsection{Personalized Ranking of Reviews}
As we already have the documents (reviews) and the query (the user profile), we pass them to  BM25 scoring function to obtain the scores of each review. We then rank the reviews in the descending order of the scores.
As an example in our experiments, when a user A1Z1LLEQQ4D1IQ visits a product B00AIQHQZS, the best ranked review and least ranked review are as follows,
\newline

Best ranked review : "I have been using this cable just under three months and it progressively did not charge my phone when plugged in. I would try unplugging it about 5-10 times before it worked continuously. And I'm not writing this review after repeat efforts to get it to work but to no avail. My original cable works perfectly so I know its not the port on the phone. The detachable USB portion saves me by still allowing me to use my original cable but its just a hassle to remember to take it with me everywhere I go because iPhone battery life sucks. Best to spend the money and get a better charger than this one."

Least ranked review : 'not working in the beginning. just too lazy to return it.'

It is evident that while both the reviews say negatively about the product, the best ranked review gives a clear overview of how the product doesn't work while the least ranked doesn't mention anything about in detail.
\newline

In our experiment, we have observed that the model automatically pushes down the short reviews which has no clear information and are not useful of any of the user. Another observation is that, while the default ranking of Amazon is primarily done by helpfulness votes followed by time of review and other factors, the review which has highest votes is ranked somewhere in middle by our model because it might be relevant to the preferences of the user.

\section{Evaluation}
Evaluating the results in review ranking is challenging since we won't have the relevance labels and the relevance also varies from user to user, so the traditional metrics cannot be employed in personalized reviews setting. Feedback from user clicks is also not available since users don't click on reviews unlike web results. All other forms of relevance feedback like position of eye are costly and not easy to implement. Hence we evaluate the results in two ways.

\subsection{Empirical Evaluation}
We scraped the Titles of around 1000 products from Amazon for the products present in the dataset. We chose 75 interesting titles as browsing history, 20 most interesting titles as shopping history and written reviews manually to 5 products. We assume that every browsed product is liked and assigned an equal weight of '1' for all. We did this through titles since it is not feasible to visit all the 1000 products and collect time statistics but titles gives some sense of importance. Using this activity we constructed the user profile and generated the review ranks on a final target product. We manually read the reviews and ranked them before looking at the ranking generated from our model and the manual ranking is taken as the ground truth. The results compare as follows :\\

Manual (ideal) Ranking \ \ \ \ \ \ \ \ \ \ 1 2 3 4 5 6 7 8 9

Original (helpful votes+time)  \  3 1 5 4 2 7 6 8 9

Our ranking method \ \ \ \ \ \ \ \ \ \ \ \ \ \ \ 2 1 3 5 4 6 9 7 8\\

If the top 3 documents are considered to be relevant, our personalized ranking method has all the 3 reviews ranked in top which is not the case with default ranking which shows the model learned the preferences and ranked the reviews well. Another observation is that the least ranked reviews in all the above 3 ranking methods are the 'obvious' bad reviews which are very short and had no information similar to the one in Section 5.2 . We did this evaluation only on 2 products since it is very time consuming, hence we did not calculate MAP and other metrics.

\subsection{Evaluation based on review scores}
Amazon already ranks the reviews from various factors, as already discussed, hence they are expected to good enough even though not personalized. With that, it's better to assess if the overhead we took to rank the reviews in a personalized manner is really worth or not. Is there really any difference between the review ranking order by Amazon and our proposed ranking model ? To answer this question, we proposed this evaluation metric named 'satisfaction score' which is based on the scores of reviews. BM25 scores the review documents when the user profile is passed as the query. These scores, by itself doesn't provide more information and are not consistent over runs for different set of documents, but we can use them to compare both ranking methods. It also helps to answer the question - "Assuming our system perfectly models user interests (needs), what would be the scope of increase in user satisfaction ?" We define user satisfaction by "Ranking Satisfaction Score (RSS)" which is the Cumulative Positive-Weighted Score.
\begin{equation}
RSS = \displaystyle\frac{\sum^{n-1}_{i=0}{[s_i * (n - i)]}}{n}
\end{equation}
where  n is number of reviews and $s_i$ is the score of review ranked at position i.\\

This means, our ranking system will have the highest RSS, as our assumption itself is that the our system perfectly models user interests. Now this is helpful to see how much increase in satisfaction we get over default ranking since in the case our system ranks the same as Amazon, the difference in RSS would be zero. We also use this weighting scheme (5,4,3,2,1 for 5 reviews) instead of something as inverse reciprocal rank (5.0, 2.5, 1.66, 1.25, 1) because user typically reads more than 1 review to make a shopping decision and inverse reciprocal rank penalizes lower ranks on very high magnitude while our weights slowly decreases the reward which suits the purpose.\\

An example from our experiments looks as follows,\\
In "default ranking" : 66897.73314318295\\
In "our ranking" : 82537.05329346986\\
Percent increase in position based cumulative score : 23.38 \% \\
We observed an average of ~20\% increase in satisfaction when evaluated for 1000 products
\section{Reviews based Product Recommendation}
We can leverage the user profile to also recommend products on the basis of a personalized product rating instead of the generic overall rating which Amazon currently has. Since we already have the terms that user cares the most, we calculate the average rating of each term in that product. This is done by picking the reviews of the product which has a particular term $t$ , along with the ratings associated with each of the review and calculating the average rating over the term $t$. This also enables us to give the rating over each term and the user can make an effective decision without even reading the reviews.
\begin{figure}[htp]
\centering
\includegraphics[width=8cm]{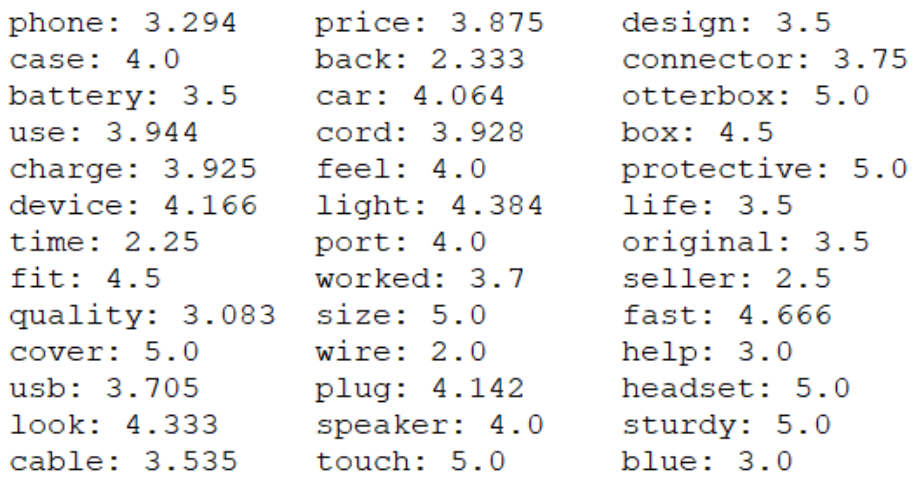}
\caption{Example of terms in user profile and their corresponding ratings}
\end{figure}

Looking at the ratings as shown in Figure. 2, the user can immediately buy the product if he cares for terms size, fit or sturdy and skip the product if he cares for seller and after purchase service. By averaging all these ratings on the top k terms, we can get the personalized recommendation score for the user on any product.

\section{Limitations \& Future work}
While the user profile learns some user behaviour which is common for all the categories (Eg - price vs quality) but to learn the preferences that are relevant to that category, the user needs to have enough browsing or shopping history. Also, most of the products don't have reviews which requires more history before the personalized recommendation is offered. As already discussed in the previous sections, the evaluation of this model is not easy due to lack of relevance feedback so to exactly use the proper weights in model, it is recommended if some elite users (having high activity) are incentivized to give relevance feedback manually, for the products they visit. This also helps to better evaluate the model with the traditional evaluation metrics and compare with the results with the baseline ranking system.

For the future work, there are several possible improvements one can work on. First would be to explore the possibility of weighted queries, since the terms in the user profile have different weighted frequencies and hence a different importance. Next, current user profile query is same for all the products in a category, and it would be helpful to re-construct the query based on the topical words according to reviews of the target product. If a user has no previous activity on a certain category, the user similarity and product similarity can be explored to borrow the profile from other similar users or profile of other category on a similar product. One can also explore the use of other scoring functions to see the performance change and define a new scoring function suitable to the reviews setting, if needed. Finally, although the weights for different types of user histories are prefixed by the current design, in the future, we would like our algorithm to able to tune the weights by learning user browsing behaviour in real time as the browsing time and purchase capacity is not same for all users. In this way, we believe it will make the ranking algorithm more flexible and adaptable to users' various information needs.

\section{Conclusion}

In this paper, we tried to enhance the customer shopping experience by incorporating personalization aspects into the review ranking process to aid customer decisions when they are making purchases on shopping websites. Specifically, we used the mobiles and accessories category of UCSD Amazon reviews dataset, and constructed the user profiles based on the simulated browsing and shopping activity for all the users. For the creation of user profiles, we came up with Review Text Frequency Model, which is a word-frequency based approach to construct user profiles hinged on the review texts of items that user buys or visits. Then we pass this user profile as query to BM25 scoring and get the personalized rankings of reviews for each user. In order to compare the performance of our ranking system to that of system currently deployed by Amazon, we conduct empirical evaluation through Ranking Satisfaction Score. Based on this, we find that the performance of our ranking method can exceed that of the default one by 20\%. Furthermore, we demonstrated the capability of user profiles to recommend products in a personalized manner and also provide ratings at terms level. Based on the experimental outcome, we validate our initial assumption that user satisfaction is expected to increase when the system ranks reviews and recommends products on a personal basis. 

\section{Acknowledgments}

We would like to thank Haoran, Rakshita, Kai and Professor Hongning Wang for their contributions and suggestions in developing this work.

\nocite{*}
\bibliography{refs}
\bibliographystyle{unsrt}

\end{document}